\documentclass[12pt,preprint]{aastex}

\slugcomment{to appear in {\it The Astrophysical Journal} on March 20, 2002}
\shorttitle{Radio-to-FIR SED for Star Forming Galaxies}
\shortauthors{Yun \& Carilli}

\begin{document}

\title{Radio-to-FIR Spectral Energy Distribution 
and Photometric Redshifts for Dusty Starburst Galaxies}

\author{Min S. Yun}
\affil{University of Massachusetts, Department of Astronomy, 
Amherst, MA~~01003}
\email{myun@astro.umass.edu}
\and
\author{C. L. Carilli}
\affil{National Radio Astronomy Observatory, P.O. Box 0, Socorro, NM~~87801}
\email{ccarilli@nrao.edu}

\begin{abstract}

As a logical next step in improving the radio-to-submm spectral
index as a redshift indicator  (Carilli \& Yun), 
we have investigated a technique of using 
the entire radio-to-FIR spectral energy distribution (SED) for 
deriving photometric redshifts for dusty starburst galaxies
at high redshift.
A dusty starburst SED template is developed from theoretical
understanding on various emission mechanisms related to
massive star formation process, and the template parameters are
selected by examining the observed properties of 23 IR selected 
starburst galaxies: $T_d=58$ K, $\beta=1.35$, and $f_{nth}=1$.
The major improvement in using this template SED for deriving
photometric redshifts is the significant reduction in 
redshift uncertainty over the spectral index technique, particularly
at higher redshifts.  Intrinsic dispersion in the radio and FIR
SEDs as well as absolute calibration and measurement errors
contribute to the overall uncertainty of the technique.
The derived photometric redshifts for five submm galaxies with known
redshifts agree well with their spectroscopic redshifts within
the estimated uncertainty.  Photometric redshifts for seven submm
galaxies without known spectroscopic redshifts
(HDF850.1, CUDSS14.1, Lockman850.1, SMM~J00266+1708,
SMM~J09429+4658, SMM~J14009+0252, FIRBACK~J1608+5418) are derived.

\end{abstract}


\section{Introduction}

Sensitive observations at submm wavelengths
are revealing what may be a population of active star forming galaxies
at high redshift which are unseen in deep optical surveys due to dust
obscuration \citep{Smail97,Barger98,Hughes98,Eales99}.
The differential source counts clearly indicate 
a large excess of far-infrared sources by a factor of 10-50 over the 
no-evolution models derived from the optical 
deep survey data \citep{guiderdoni98,Blain99a,Matsuhara00,Scott01},
and a large fraction of star formation may be hidden by dust. 
The analysis of the submm SCUBA source counts and inferred
redshift distribution based on the radio-submm flux density 
ratio \citep{Carilli99} suggests that luminous dusty galaxies 
may dominate the star formation history at early epochs
($z\sim 1-3$) \citep{barger00}. 
Many of these faint submm sources are found to be
very faint, red ($R\ge25, K\ge21$) sources \citep{Smail99a},
whose optical redshifts may be inaccessible
even for the 10-m class telescopes.  These galaxies are 
nearly entirely missing from the optical studies of star
formation at high redshifts \citep [e.g.][]{Steidel99, 
Chapman00}, and a significant revision to the optically
derived cosmic star formation history may be needed 
\citep [see][]{Blain99a}.

Because direct redshift measurements are not possible in most cases,
the redshift distribution and the cosmic evolution of the dusty
submm galaxy population are not well determined at the moment.
In a recent paper, we have proposed a technique of using the
radio-to-submm spectral index as a redshift indicator. 
This technique is based on the universal radio-to-far infrared
(FIR) correlation for star forming galaxies \citep{Condon92},
with the assumption that the spectral shapes may be 
similar enough to be able to differentiate between low and high
redshift objects \citep [see][]{Carilli99}. 
To understand the magnitude of scatter in this relation and
resulting uncertainty in the redshift estimates,
we derived an empirical $\alpha^{350}_{1.4} - z$ 
relation\footnote{$\alpha^{350}_{1.4}$ is the spectral index
between 1.4 GHz and 350 GHz.} using
the observed spectral energy distributions (SEDs) of 
17 low redshift starburst galaxies (Carilli \& Yun 2000; also
see Dunne, Clements, \& Eales 2000a).  There is 
a significant scatter, but the existing data on high redshift
star forming galaxies and AGN-hosts appear to follow this 
relation well.

While the radio-to-submm spectral index has been demonstrated
to be a useful redshift indicator, using it to derive a redshift for
any particular object may be risky because of the scatter
in the observed $\alpha^{350}_{1.4} - z$
relation and the flattening in the relation at high redshifts.  
Obtaining redshift estimates using flux density ratios at other
wavelength bands has also been suggested although none is
particularly more successful \citep[e.g.][]{Hughes98,Blain99b,Hines99,Fox01}.
More accurate redshift estimates for dusty 
starbursts may be obtained by utilizing the information associated
with the {\it entire radio-to-FIR SED}.  
Such a photometric redshift technique requires a good SED template,  
and here we derive an SED template based on the theoretical 
expectations of thermal dust, thermal Bremsstrahlung, and 
non-thermal synchrotron emission from a dusty starburst galaxy.
Our SED template is constructed by examining the properties of 
the 23 IR selected starburst galaxies whose submm/FIR SED data are 
readily available in the literature.
When applied to a sample of submm galaxies with known
redshifts, the resulting redshift estimates are in excellent
agreement with the spectroscopic redshifts.  We also report 
photometric redshifts for a sample of submm galaxies without
known spectroscopic redshifts, including the brightest 
SCUBA source in the Hubble Deep Field \citep{Hughes98}. 

%The organization of the paper is as follows.  
%The radio-to-FIR SED model is described in \S2, largely following
%the discussion of star forming galaxies by \citet{Condon92}.
%A dusty starburst SED template is also derived from the ensemble average  
%of 23 IR selected starburst galaxies.  A
%photometric redshift technique utilizing ths dusty starburst SED
%is tested on a sample
%of submm selected galaxies with known redshifts, 
%and this method is extended to a small sample of submm galaxies 
%without known redshifts drawn from the literature in \S3.
%Uncertainties of our photometric redshift technique resulting
%from various physical effects are discussed and evaluated in \S4.
%Finally, the summary and implications of this work discussed in \S6. 

\section{Starburst SED Model \label{sec:model}}

The observed SED from radio to infrared wavelengths for a dusty 
starburst galaxy is dominated by the energetic signatures of 
young stars and their environment.  The bulk of 
the radiation from young OB stars is absorbed by dust 
and re-emitted in the infrared.  The HII regions surrounding the 
young stars are also the sources of thermal Bremsstrahlung 
(free-free) emission.  Energetic electrons produced in supernovae
and their remnants are thought to account for nearly all of the 
radio synchrotron radiation. Following the discussion by  
Condon (1992), a simple starburst SED model is constructed as a
linear sum of dust continuum ($S_{d}$), thermal Bremsstrahlung or
free-free ($S_{th}$), and non-thermal synchrotron ($S_{nth}$) emission.

%We make a simple physical assumption that a star forming
%galaxy consists of a collection of discrete star forming
%regions or clouds, which may be clustered and concentrated
%at one location or distributed over a large area.  
%Then the observed SED only depends on the number of these
%clouds and thus the total star formation rate.

%The variations in 
%physical properties such as dust temperature or cosmic ray 
%production efficiency can affect the.  

\subsection{Thermal Dust Emission \label{sec:Sdust}}

The flux density, $S_d(\nu)$, from a cloud at a distance
$D_L$ containing $N$ spherical dust grains each of cross
section $\sigma_d$, temperature $T_d$, and emissivity $Q(\nu)$,
is given by
\begin{equation}
S_d(\nu) = N(\sigma_d /D_L^2)Q(\nu)B(\nu,T_d), \label{eq:Sd1}
\end{equation}
\noindent
where $B(\nu,T_d)$ is the Planck function describing  
thermal black body radiation.  The term $N(\sigma_d /D_L^2)$
is the solid angle $\Omega_d$ subtended by the dust
source in the sky.  For the emissivity function,
we adopt a form 
\begin{equation}
Q(\nu) = 1 - exp~[ - ({{\nu}\over{\nu_c}})^\beta ], \label{eq:Q}
\end{equation}

\noindent
where $\beta$ is the dust emissivity index generally thought
to be between 0 and 2 \citep{Hildebrand83}.  
This particular functional form allows
the dust spectrum to be that of a purely black body above the
critical frequency $\nu_c$ where the dust clouds become 
optically thick while it is that of the $\nu^{2+\beta}$
grey body spectrum at lower frequencies.   Observations suggest that
thermal dust emission is nearly optically thick even at 
$\lambda \sim 100$ $\mu$m for galaxies dominated by intense 
starburst regions \citep [see][]{Scoville91,Solomon97}, and
we adopt $\nu_c = 2 \times 10^{12}$ Hz 
($\lambda_c = 150$ $\mu$m).  Combining the 
Eqs.~\ref{eq:Sd1}~\&~\ref{eq:Q}, 
\begin{equation}
S_d(\nu) = \Omega_d B(\nu,T_d)(1-e^{-({{\nu}\over{\nu_c}})^\beta}). 
\label{eq:Sd2}
\end{equation}
\noindent For a dust source of $\theta$ arcsec in diameter, expected 
flux density at $\nu$ (in GHz) due to thermal dust emission is
\begin{equation}
S_d(\nu) = 2.8\times 10^{-8}  
{{\nu^3 {\theta}^2}\over{e^{[0.048\nu/T_d]}-1}}
(1-e^{-({{\nu}\over{2000}})^\beta}) ~~{\rm Jy}. 
\label{eq:Sd3}
\end{equation}

The FIR luminosity, $L_{FIR}$, is derived by integrating Eq.~\ref{eq:Sd3}
over the emitting area $\Omega_d$ and over the frequency range
corresponding to $\lambda = 40-500~\mu$m,
$L_{FIR}=4\pi D_L^2 \int \int S_d(\nu) d\nu d\Omega$.
The massive star formation rate (SFR) for a dusty starburst galaxy 
is then directly related to its dust continuum spectrum by 
the FIR luminosity \citep [i.e. $SFR=L_{FIR}/(5.8\times 
10^9 L_\odot)~M_\odot~{\rm yr}^{-1}$,][]{Kennicutt98}

\subsection{Thermal Bremsstrahlung (free-free) Emission 
\label{sec:Sff}}

Thermal Bremsstrahlung (free-free) emission from  
ionized gas in the HII regions surrounding hot young stars 
can be described in the same functional form as the
thermal dust emission as in Eq.~\ref{eq:Sd2},
\begin{equation}
S_{ff}(\nu) = \Omega_{ff} B(\nu,T_e)(1-e^{-\tau_{ff}}), 
\label{eq:Sff1}
\end{equation}
\noindent where $T_e$ is electron temperature and $\tau_{ff}$
is free-free optical depth at the observed frequency.  
In the Rayleigh-Jeans regime ($h\nu\ll kT$), 
$B(\nu,T_e)=2kT_e\nu^2/c^2$ at radio frequencies, and
\begin{equation}
S_{ff}(\nu) = 2kT_e\nu^2 c^{-2} \Omega_{ff} (1-e^{-\tau_{ff}}). 
\label{eq:Sff2}
\end{equation}

Free-free optical depth is given by
$\tau_{ff} = 0.083 EM \nu^{-2.1} T_e^{-1.35}$, where
$EM \equiv \int n_e^2 dl$ (cm$^{-6}$ pc) is the emission 
measure.  In the optically thin regime, thermal 
Bremsstrahlung emission then has the familiar $\nu^{-0.1}$ dependence.
A circular source of $\theta$ arcsec in diameter would have flux density of
\begin{equation}
S_{ff}(\nu) = 4.8 \times 10^{-8} EM \nu^{-0.1} 
T_e^{-0.35}{\theta}^2 ~~{\rm Jy}.
\label{eq:frfr2}
\end{equation}

\subsection{Non-thermal Synchrotron Emission \label{sec:Snth}}

For an ensemble of relativistic electrons with a power-law 
distribution of energy $N(E)=N_\circ E^{-\gamma}$ and an 
isotropic velocity distribution, their synchrotron emission
coefficient is  
\begin{equation}
\epsilon \propto N_\circ (B sin \theta)^{(\gamma+1)/2} \nu^{(1-\gamma)/2},
\label{eq:eps1}
\end{equation}
\noindent where $\theta$ is the angle between the magnetic field $B$ 
and the line of sight to the observer.  The resulting power-law
spectrum has a form $S_\nu \propto \nu^{-\alpha}$ where $\alpha =
(\gamma -1)/2$ is the spectral index. \citet{Bell78} proposed the
acceleration of cosmic rays by supersonic shocks (of velocity
$v_s$) as the main mechanism
for non-thermal synchrotron emission from supernova remnants
and derived the volume emissivity as,
\begin{equation}
\epsilon(\nu) \sim 3\times 10^{-33} \nu^{-\alpha}({{n}\over{cm^{-3}}})
({{\alpha}\over{0.75}})({{v_s}\over{10^4km/s}})^{4\alpha}
({{B}\over{10^{-4}G}})^{\alpha+1} 
~~{\rm erg~s^{-1} Hz^{-1} cm^{-3}}
\label{eq:eps2}
\end{equation}

Assuming supernova remnants dominate the non-thermal synchrotron 
emission from galaxies, \citet{Condon92} derived a relation 
$S_{nth}\propto \nu_{SN}$ where $\nu_{SN}$ is the Type II 
supernova rate per year (see his Eq.~17).  However, this relation 
predicts a $\nu_{SN}$ more than an order of magnitude too
large for our Galaxy, and Condon proposed a revised relation 
based on Galactic normalization (i.e. his Eq.~18), which 
can be re-written as
\begin{equation}
S_{nth}(\nu) = 1150~\nu^{-\alpha} \nu_{SN} {D_L}^{-2}~{\rm Jy},
\label{eq:nth1}
\end{equation}
where $D_L$ is luminosity distance in Mpc.

\subsection{Starburst SED Template \label{sec:template}}

The first and the foremost important step in all photometric redshift
technique is establishing a robust and reliable template.
Key SED template parameters for the starburst SED template are 
$T_d$ and $\beta$ 
for the dust emission (Eq.~\ref{eq:Sd3}), $T_e$ and $EM$ for thermal 
free-free emission (Eq.~\ref{eq:frfr2}), and $\nu_{SN}$ and $\alpha$ 
for the non-thermal synchrotron emission (Eq.~\ref{eq:nth1}).
These parameters and their dispersions 
are determined by deriving best fit SED models for the 23 IR selected 
starburst galaxies whose submm/FIR data are readily 
available in the literature.  

Flux density for dust emission $S_d$ can be derived directly 
using Eq.~\ref{eq:Sd2} by specifying the source 
size $\Omega_d$, dust temperature $T_d$, and emissivity $\beta$.  
The effective source solid angle $\Omega_d$ for a given source
is derived from the observed 100 $\mu$m flux density, assuming dust 
emission is nearly optically thick (see \S~\ref{sec:Sdust}).  
For example, the prototypical ultraluminous 
galaxy Arp~220 is a 114 Jy ($1.14 \times 10^{-21}$ erg cm$^{-2}$ 
Hz$^{-1}$) source at 100 $\mu$m, which translates to 
$\Omega_d = 3 \times 10^{-11}$ $rad^2$ for $T_d=59$ K 
(see Table~\ref{tab:23gals}).  
This corresponds to a 
circular area with 1.2$''$ (390 pc) in diameter\footnote{An ensemble 
of distributed sources with the equivalent total area is also allowed 
by this simple model.}, which agrees well 
with the size of the molecular gas complex fueling the 
nuclear starburst traced in CO \citep{Sco97,Downes98,Sakamoto98}.
The best fit dust SED model is then selected by examining the parameter 
space for $T_d$ and $\beta$, between 25-85 K and 1.0-2.0, respectively.

Thermal Bremsstrahlung or free-free flux densities can 
be derived from Eq.~\ref{eq:frfr2} by determining the frequency 
where free-free opacity $\tau_{ff}$ becomes unity and estimating the 
size of the emitting region.  Alternatively, one can also compute 
the free-free flux density from the inferred  
SFR as it is proportional to the production rate of 
Lyman continuum photons.  Using the H$\alpha$ normalization of 
\citet{Kennicutt98} for a Salpeter IMF and assuming 50\% of Lyman 
continuum photons are quenched by dust absorption, 
Eq.~23 of \citet{Condon92} can be re-written as
\begin{equation}
L_{ff} (\nu) = 7.9\times 10^{26} \nu^{-0.1} 
({{SFR}\over{M_\odot yr^{-1}}}) ~~{\rm erg~s}^{-1}
{\rm Hz}^{-1}.  \label{eq:Condon23}
\end{equation}
\noindent The free-free flux density for a galaxy with 
a given FIR derived star formation rate is then
\begin{equation}
S_{ff}(\nu) = 0.71~\nu^{-0.1}({{SFR}\over{M_\odot yr^{-1}}}) 
{D_L}^{-2}~~{\rm Jy}.
\label{eq:Sff}
\end{equation}
\noindent The free-free flux density $S_{ff}$ determined this way is 
nearly identical to the value derived using Eq.~\ref{eq:frfr2}
assuming $\tau_{ff}\sim 1$ at $\nu=1$ GHz and $\Omega_{ff}= \Omega_d$.

Non-thermal synchrotron flux density $S_{nth}$ can also be 
parameterized as a function of SFR since $\nu_{SN} \propto SFR$.
Adjusting Eq.~20 of \citet{Condon92} for the Salpeter IMF, 
Eq.~\ref{eq:nth1} can be re-written as,
\begin{equation}
S_{nth}(\nu) = 25~f_{nth} \nu^{-\alpha}
({{SFR}\over{M_\odot yr^{-1}}}) {D_L}^{-2}~~{\rm Jy}.
\label{eq:nth2}
\end{equation}
\noindent The synchrotron spectral index $\alpha$ is known to 
lie within a narrow range around
0.7-0.8, and we simply adopt $\alpha=0.75$ to minimize the number 
of free parameters in our model.  
Condon's original derivation arbitrarily employed a 
Galactic normalization as noted earlier.  We thus add a  
scaling factor $f_{nth}$ (of order unity) here in order to determine the 
normalization more suitable for starburst galaxies.  

Only the SED data at frequencies near and below the dust 
peak are used for the model fit because the mid- to 
near-IR data points clearly require additional, higher 
temperature dust components.
%\footnote{Mass associated with the higher temperature 
%components is miniscule as emissivity goes up as $T^{4+\beta}$.}.
The SED fit for Arp~220 is shown in Figure~\ref{fig:a220sed}, and
it demonstrates the effects of 
changing $T_d$ while holding $\beta$ constant. 
The scatter of data points about various models suggests that there are 
systematic scaling differences among different flux density measurements, 
generally larger than the nominal uncertainties reported, especially 
in the submillimeter wavelengths.  At a glance, all three dust 
temperature models plotted seem to do an adequate job of fitting 
the submm and FIR data points qualitatively.  However, a closer 
examination reveals that the $T_d=74$ K model fails to 
match the submm points.  The $T_d=43$ K model matches the submm 
data points reasonably well, but it clearly falls short on 
predicting the FIR data points.
Previous studies of submm dust properties for luminous infrared galaxies
generally favored cold dust temperature between 40 and 50 K for Arp~220
\citep [e.g.][]{Eales89,Scoville91,Dunne00a}.
If we were to choose a {\it single temperature}
dust SED model that best fits all observed data points between 1mm and
60 $\mu$m, however, a slightly warmer dust model is favored. 

\begin{figure}
\epsscale{0.75}
\plotone{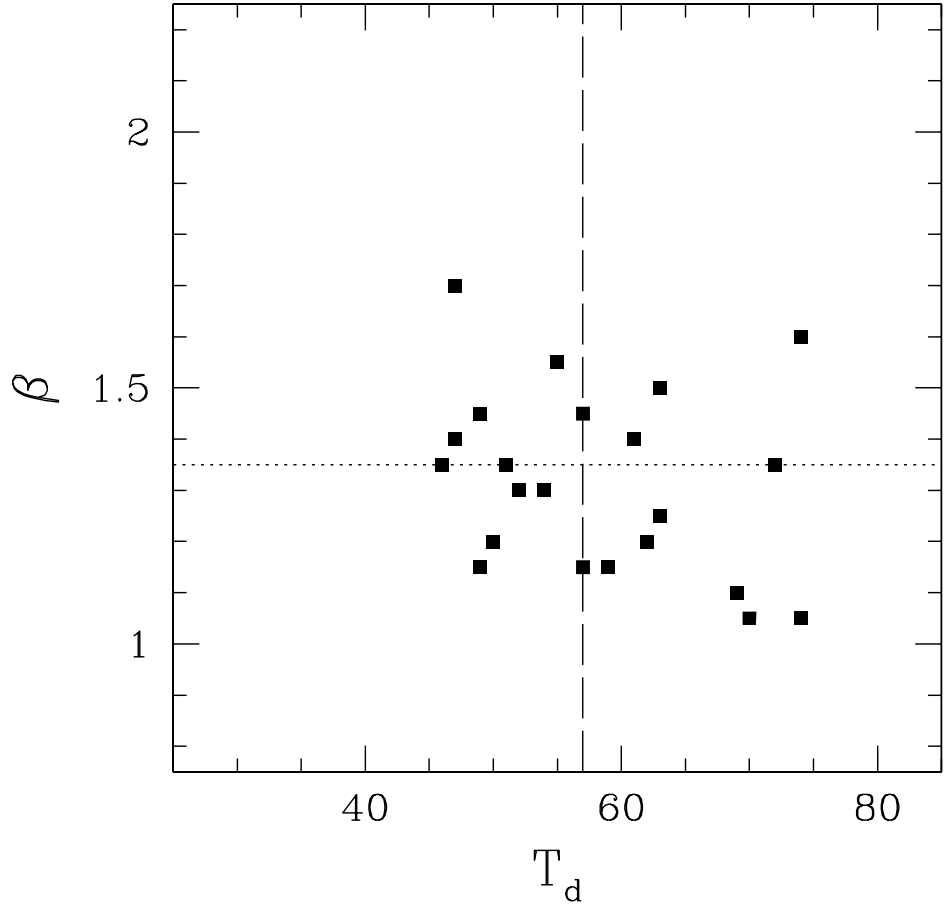}
\caption{A distribution of best fit $\beta$ values versus
$T_d$ for the 23 luminous IR starburst galaxies.  The median values
for $\beta$ and $T_d$ are 1.35 \& 57 K while the mean values are
$1.32\pm0.17$ and $58\pm9$ K, respectively.
\label{fig:betaTd}}
\end{figure}

Our starburst SED model analysis is applied to 23 IR-selected 
starburst galaxies that are selected for  (1) their
$L_{FIR}$ exceeding $10^{11}~L_\odot$; and (2) having at least two 
measurements covering their submm part of the spectrum. 
The SED data are primarily drawn from the IRAS Faint Source 
Catalog, NRAO/VLA Sky Survey \citep{Con98}, \citet{Con91b},
\citet{Carico92}, \citet{Rigopoulou96}, \citet{Benford}, 
\citet{Lisenfeld99}, \citet{Dunne00a}, and \citet{Dunne01}.  As summarized
in Table~\ref{tab:23gals}, the characteristic dust 
temperature $T_d$ ranges between 46 K and 74 K, with a mean of
$58\pm9$ K.  Dust emissivity $\beta$ also ranges widely, between
1.05 and 1.70, with a mean of $1.32\pm0.17$ and a median of 1.35
(see Figure~\ref{fig:betaTd}).  
It is likely that all galaxies consist of an ensemble of dust clouds 
with a range of temperature \citep[see][]{Yun98,Frayer99,Dunne01}, 
and single temperature SED models tend to favor a smaller $\beta$ 
(flatter submm dust spectrum) and higher dust temperature than 
typically found in giant molecular clouds (10-20 K).  
Since available SED measurements are generally sparse in practice, 
we accept these limitations of a single temperature dust model for 
the purpose of keeping the number of free parameters manageably small.

\begin{figure}
\epsscale{0.75}
\plotone{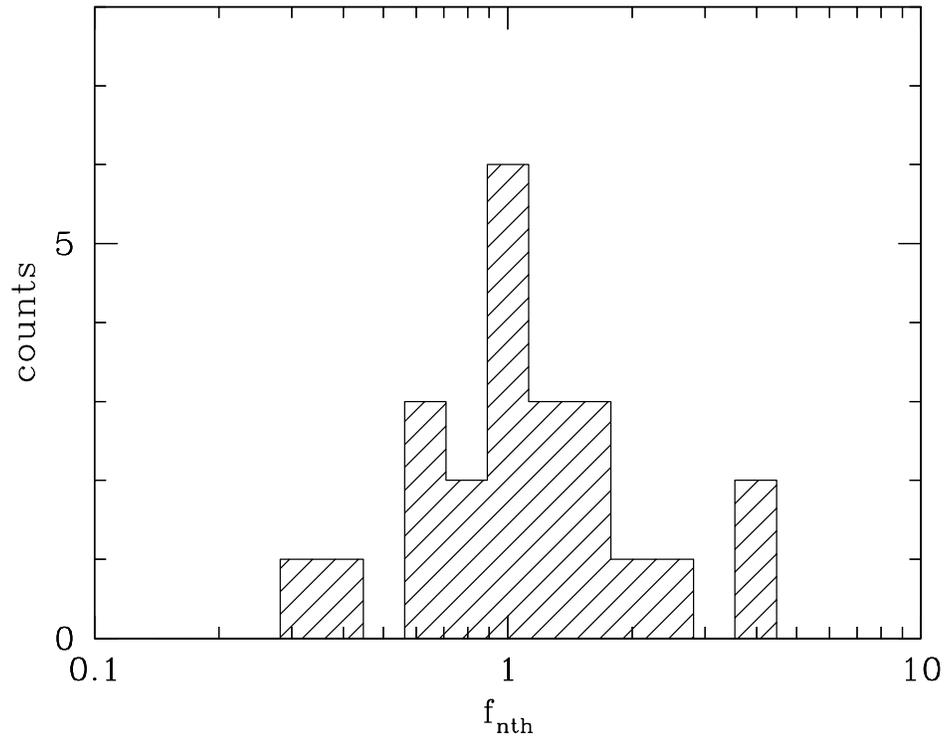}
\caption{A distribution of $f_{nth}$ values for the 23 luminous 
IR starburst galaxies.  
%There is well over an order of magnitude spread in $f_{nth}$,
%suggesting some variation among galaxies in converting radio luminosity
%to massive star formation rate.  
The median for the whole sample is 1.1, and it is reduced to 1.0 
if two clear radio AGN hosts are removed.
\label{fig:fnth}}
\end{figure}

Thermal Bremsstrahlung (free-free) emission makes significant
contribution only in the bottom of the SED trough between the
non-thermal synchrotron and thermal dust feature, and it plays
essentially no role in defining the starburst SED template.
Variations in the non-thermal synchrotron emission are tracked
by a parameter $f_{nth}$ (see Eq.~\ref{eq:nth2}).  The dispersion
in the radio-FIR correlation is measured to be about 0.25 in a
logarithmic scale \citep [e.g.][]{Yun01}, and a
rather broad range of $f_{nth}$ found
(see Figure~\ref{fig:fnth}) is consistent with this expectation.  Presence of 
a radio AGN (e.g. Mrk~231, NGC~6240) can account for the increase 
in some cases, but synchrotron emission efficiency may indeed vary 
from galaxy to galaxy.  The median for the whole sample
is 1.1, but it is reduced to 1.0 if the two clear radio AGN hosts
are removed.  Therefore, on average, the 
non-thermal synchrotron emission from these dusty starbursts appear to 
follow the Galactic normalization adopted by \citet{Condon92}.

In summary, we adopt $T_d=58$ K, $\beta=1.35$, and $f_{nth}=1.0$ for 
our photometric redshift SED template.
This relatively flat $\beta$ is in good agreement with the analysis 
of 102 galaxies by \citet{Dunne00a}, but the mean dust temperature 
we adopt is significantly warmer than the mean of 
$T_d = 36 \pm 5$ K Dunne et al. derived using 60 $\mu$, 100 $\mu$, and 
850 $\mu$ measurements.  The main difference is that the dust
emission from our IR selected, luminous dusty starbursts are on 
average significantly warmer than those of the Dunne et al. sample, 
which includes a large number of late type field galaxies.
%Warmer dust temperature also lowers the required dust
%mass for a 10 mJy 850 $\mu$m source at to about $10^8 M_\odot$ 
%while a colder dust (20-30 K) source would require $10^9 M_\odot$,
%far in excess of the typical dust mass for an $L_*$ galaxy.
%Dusty galaxies at high redshifts detected by the current 
%generation of mm/submm bolometers have FIR luminosity exceeding 
%$10^{12} L_\odot$, and our choice of a higher dust temperature 
%model is justified.

\bigskip
\section{Photometric Redshifts for Dusty Galaxies at High Redshift
\label{sec:zfit}}

\subsection{Procedure}

Once the dusty starburst SED template is chosen, then the entire
radio-to-FIR continuum spectrum for any given galaxy can be 
described in terms of just the total star formation rate 
$SFR$ and the luminosity distance $D_L$.  The dust spectrum
$S_d$ in Eq.~\ref{eq:Sd3} can be re-written as
\begin{equation}
S_d(\nu) = 1.5\times 10^{-6} 
{{SFR (1+z) \nu^3}\over{D_L^2(e^{0.00083\nu}-1)}}
(1-e^{-({{\nu}\over{2000}})^{1.35}}) ~~{\rm Jy}. 
\label{eq:Sd4}
\end{equation}
Combined with the expressions for free-free and non-thermal synchrotron 
emission given in Eqs.~\ref{eq:Sff}~\&~\ref{eq:nth2}, the 
entire radio-to-FIR continuum spectrum can be written as,
\begin{equation}
S(\nu_{obs}) = [25~f_{nth} \nu_\circ^{-\alpha}+0.71~\nu_\circ^{-0.1} +1.5\times 10^{-6} 
{{\nu_\circ^3(1-e^{-({{\nu_\circ}\over{2000}})^{1.35}})}\over{e^{0.00083\nu_\circ}-1}}]
\ {{(1+z)SFR}\over{D_L^2}}~~{\rm Jy}, 
\label{eq:SdSFR}
\end{equation}
\noindent where $SFR$ is in $M_\odot$ yr$^{-1}$, $D_L$ in Mpc\footnote{
We adopt $H_\circ=75$ km sec$^{-1}$ Mpc$^{-1}$ and $q_\circ=0.5$ for 
this paper.}, 
and $\nu_\circ=(1+z)\nu_{obs}$ in GHz.  A $(1+z)$ term is needed for
sources at cosmological distances in order to account for the 
frequency or wavelength folding by the Doppler effect.

The redshift $z$ and the FIR luminosity ($SFR$) of a 
particular galaxy can be determined simultaneously by comparing the observed 
SED with the dusty starburst template in the ($z$,$SFR$) space.
To be precise, one needs to include a correction term accounting for the  
cosmic microwave background (CMB) because most SED data
are measured in contrast to the CMB.  In practice this correction can be 
safely ignored as long as $T_d/(1+z) \gg 2.7$ K.

As a galaxy is placed further and further away,
its entire SED shifts to the bottom (fainter, because of $D_L^{-2}$) 
and to the left (lower frequency, $\nu_{obs}=\nu_o/(1+z)$) 
in the ($z$,$SFR$) space.  
This generic behavior for a dusty starburst and 
the resulting change in the apparent spectral index between 1.4 GHz and 
850 $\mu$m has been pointed out previously as a redshift indicator 
by \citet{Carilli99}.  The slope of the rising part of the dust 
spectrum is such that the Doppler shift of the spectrum nearly 
offsets the $D_L^{-2}$ drop in flux density, making the submm bands 
particularly attractive for blind searches, but the SED 
measurements on this part of the dust spectrum offer limited 
redshift information for the same reason \citep[e.g.][]{Hughes98,Fox01}.  
As in other photometric redshift techniques, the redshift information 
comes from distinct spectral features such as the trough between 
the declining non-thermal synchrotron emission in radio and the 
sharp rise in the dust spectrum or the dust peak near the rest wavelength 
of 100 $\mu$m.  Even when the dust peak in the FIR is not 
sampled by observations, the radio synchrotron measurements help set 
the vertical scale with respect to the dust spectrum, i.e. the $SFR$.

Because SED measurements for most submm galaxies include only 
a few discrete points rather than a continuous frequency coverage, 
our best fit SED model search utilizes a $\chi^2$ minimization with 
discrete sampling of the 
parameter space rather than a full cross-correlation 
technique.  One advantage of this approach is that the upper limits 
in flux density can be incorporated in a straightforward way 
by simply rejecting all trial 
SEDs that are incompatible with the upper limits.

\subsection{Trials on Galaxies of Known Redshifts 
\label{sec:tests}}

\begin{figure}
\vbox to3in{\rule{0pt}{3in}}
\includegraphics{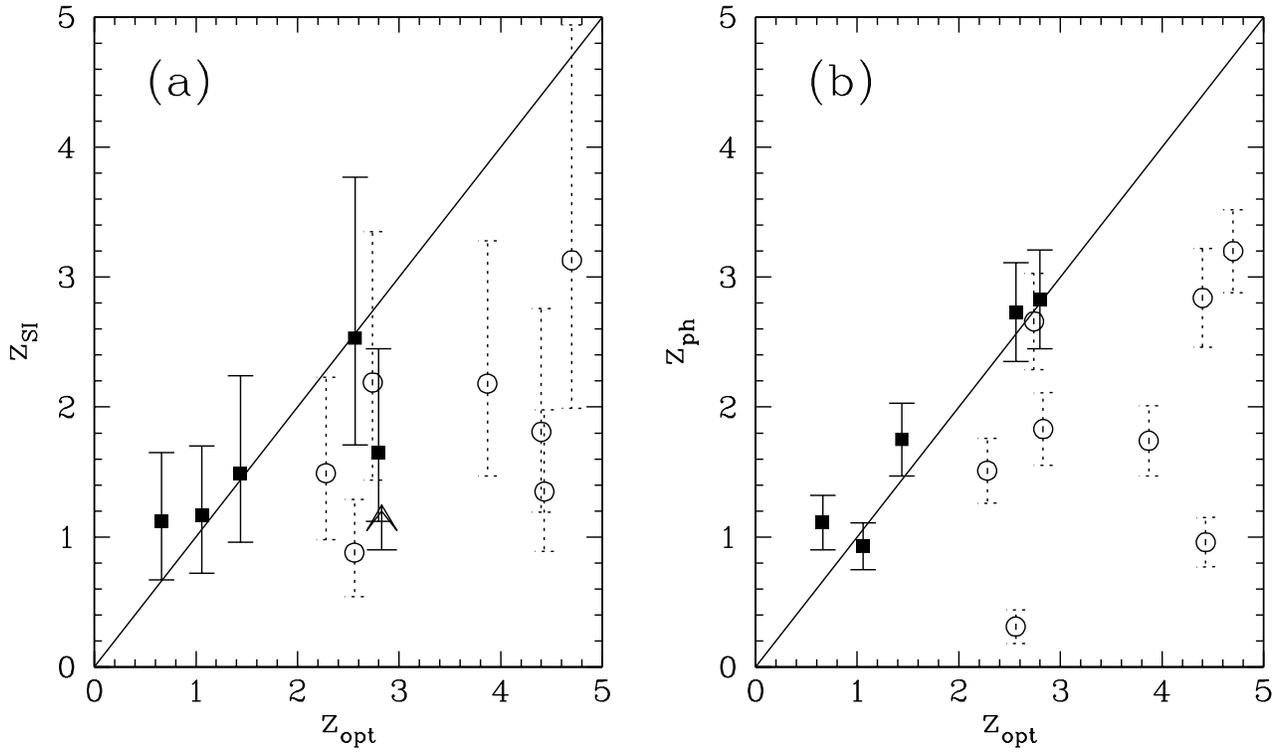}
\caption{(a) A comparison plot of spectroscopic redshifts ($z_{opt}$) versus
redshift estimates from the radio-to-submm spectral index  ($z_{SI}$). 
(b) A comparison plot of spectroscopic redshifts ($z_{opt}$) versus
radio-to-FIR SED photometric redshifts ($z_{ph}$).  Solid squares
represent submm galaxies while the empty circles are submm detected
optical QSOs.
\label{fig:comparez}}
\end{figure}

To test the robustness of our SED template, we apply this
photometric redshift technique to several well studied 
submm galaxies, and these results are summarized in 
Table~\ref{tab:ztable}.  Disregarding the formal uncertainties
for the moment, the new photometric redshifts $z_{ph}$ are
in excellent agreement with the spectroscopic redshifts $z_{sp}$
for the all five submm galaxies with known redshifts.   
When compared to the old radio-to-submm
spectral index estimates $z_{SI}$ as shown in Figure~\ref{fig:comparez},
the improvement is seen mainly at high redshift ($z>2$)
where the effectiveness of the spectral index method diminishes
due to the flattening of the $\alpha-z$ relation.
Some improvement is generally expected for the new photometric
method since more information
is utilized.  At the same time this comparison also highlights the 
efficiency of the spectral index technique which utilizes
rather limited amount of information.  
  
\begin{figure}
\vbox to3in{\rule{0pt}{3in}}
\includegraphics{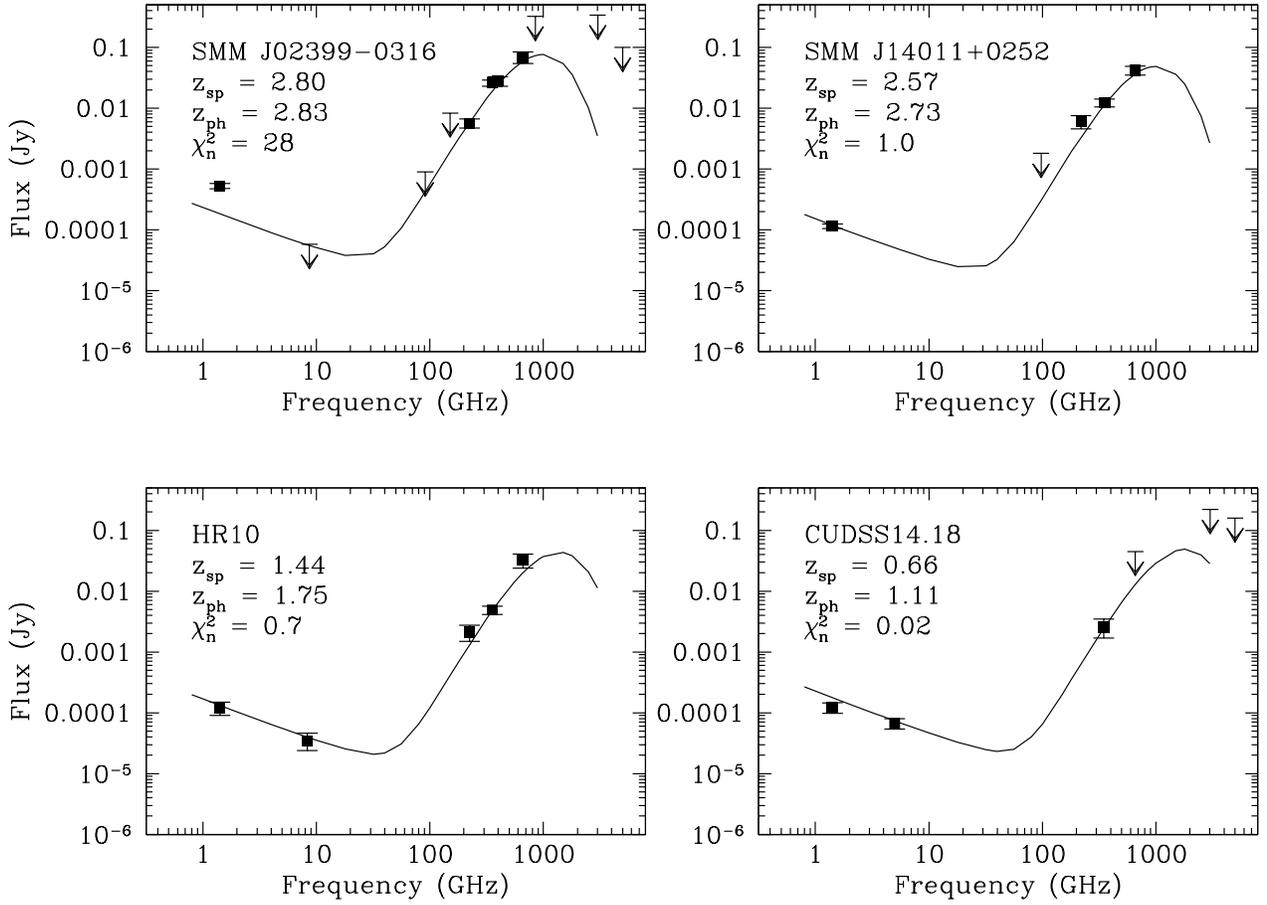}
\caption{Results of the SED fits for four submm galaxies with
known spectroscopic redshifts.
\label{fig:4seds1}}
\end{figure}

The best fit models are shown in Figure~\ref{fig:4seds1}
for the four cases for which the most SED data exist.
For the two well studied SCUBA sources SMM~J02399$-$0136 
\citep [$z=2.80$,][]{Ivison98} and SMM~J14011+0252 
\citep [$z=2.57$,][]{Ivison00}, we derive photometric redshifts of
2.83 \& 2.73, respectively.  The agreement with 
their spectroscopic redshifts is good in both cases, 
and the improvement over the 
$\alpha-z$ estimate is quite substantial for 
SMM~J02399$-$0136.  The excellent agreement in SMM~J02399$-$0136
is somewhat fortuitous since the 1.4 GHz measurement is well above
the best fit SED model (thus the source of a large reduced $\chi^2$), 
but the best fit solution is limited 
correctly by the 3$\sigma$ upper limit at 8.7 GHz.  These two
radio measurements are not consistent, and either the radio AGN
is highly variable or the 1.4 GHz measurement may be affected by 
source confusion.  The photometric redshifts ($z_{ph}$) for the submm 
galaxies SMM~J02399$-$0134 \citep [$z=1.06$,][]{Smail00} and HR10 
\citep [$z=1.44$,][]{Cimatti98,Dey99} are also in good agreement with 
their spectroscopic redshifts within the uncertainty associated with 
the dust SED parameters (see \S~\ref{sec:err-total}).  
The SED fit for HR10 shown in 
Figure~\ref{fig:4seds1} suggests that the radio continuum
is fainter than expected (i.e. $f_{nth}<1$), which 
drove the $z_{ph}$ to a slightly higher value.  
Only in CUDSS14.18 \citep [$z=0.66$,][]{Lilly99} does the photometric 
redshift differ from the spectroscopic redshift by an 
amount larger than the nominal uncertainty of the technique.  
Only one submm SED measurement exists for this galaxy, and this 
may be an important limiting factor.  CUDSS14.18 also appears to be   
slightly underluminous in radio continuum.

\subsection{Photometric Redshifts for Submm Galaxies
\label{sec:estimates}}

\begin{figure}
\epsscale{0.90}
\plotone{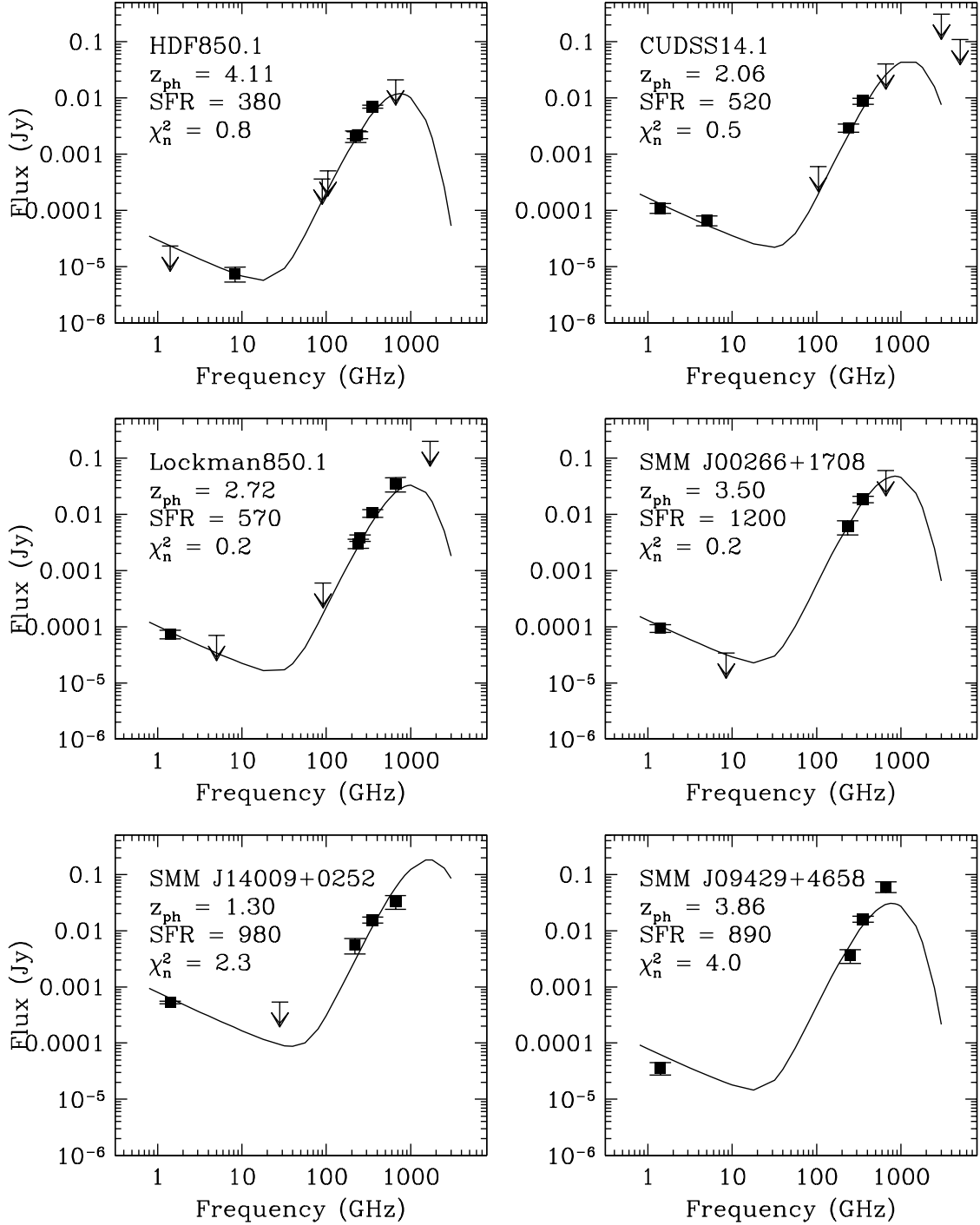}
\caption{Results of the SED fits for six submm galaxies with
unknown redshifts are shown, including HDF850.1.
\label{fig:4seds2}}
\end{figure}

Since the great majority of the submm galaxies are too faint to yield
optical redshifts, a photometric technique may be the best way
to infer their redshifts and evolution. We apply our photometric 
redshift technique to seven submm galaxies whose spectroscopic redshifts 
are unknown but are suspected of being at high redshifts: HDF850.1, 
Lockman850.1, CUDSS14.1, FIRBACK~J1608+5418, SMM~J00266+1708, SMM~J09429+4658,
SMM~J14009+0252.  The photometric redshifts and $SFR$ 
we derive are summarized in 
Table~\ref{tab:ztable}, and the best fit models  
for six cases are shown in Figure~\ref{fig:4seds2}.

Perhaps the most interesting case is
HDF850.1, which is the brightest SCUBA source in the Hubble Deep Field 
\citep{Hughes98,Downes99a}.  Despite the extraordinarily deep HST 
imaging data available, this submm galaxy has eluded optical
identification entirely.
From the radio-to-submm flux density ratio, we have previously estimated 
its redshift to be $z>2.6$, but the flattening 
of the $\alpha-z$ relation at $z>2$ meant that its redshift was
not very well constrained \citep{Carilli99}.  
Our new photometric redshift analysis suggests that this galaxy 
has a high likelihood ($\chi^2_n \sim 1$) of 
having a redshift near $z\sim 4.1$ with an intrinsic IR 
luminosity of about twice that of Arp~220.  The uncertainty 
in the inferred redshift is too large ($\sigma_z \sim 0.5$) for 
a CO search using existing 
instruments, but a spectroscopic verification of HDF850.1 and
other optically faint submm galaxies should become possible 
using future broadband instruments on the Green Bank Telescope 
or the Large Millimeter Telescope.

SMM~J09429+4658 is another historically significant
SCUBA galaxy, whose extremely dusty nature and red color
($K=19.4$ mag, $I-K>6$) was clearly 
demonstrated for the first time by the combined analysis of the 
high angular resolution VLA continuum 
imaging and deep near-IR imaging \citep{Ivison00}. Our new photometric 
redshift analysis suggests a moderately
high redshift ($z_{ph} \sim 3.9$), but the SED model fit  
(Figure~\ref{fig:4seds2}) is quite poor ($\chi^2_n \sim 4$) as our 
SED template cannot be fully reconciled with the three existing
SED measurements.  One possible explanation is that this submm
galaxy is underluminous in radio continuum as in HR10 and CUDSS14.18, 
and its actual redshift is smaller.

SMM~J00266+1708 is another clear example of an optically 
faint ($K=22.5$ mag)
and red submm galaxy, whose extreme properties are
clearly established by the combined analysis of the high angular
resolution millimeter continuum observations at OVRO and 
deep near-IR imaging using the Keck telescope \citep{Frayer00}.
The derived photometric redshift of $z_{ph}=3.50$ makes it
one of the highest redshift objects examined in this study,
perhaps only the second after HDF850.1, and it has the second
largest intrinsic FIR luminosity after SMM~02399$-$0136.

CUDSS14.1 is the brightest SCUBA source discovered in the CFRS 14hr 
field by \citet{Eales00}.  Its optical counterpart has been identified 
using high angular resolution imaging at the VLA and IRAM interferometer 
\citep{Eales00,Gear00}, but it has not yet yielded a spectroscopic 
redshift.  All four significant detection
points lie along the model SED while all upper limits are 
consistent with the model.  The derived photometric redshift of
$z_{ph}=2.06$ is in good agreement with its optical photometric 
redshift of $z\sim 2$ \citep{Lilly99}.  Using the SED model of 
\citet{Dunne00a} and optical to near-IR color, \citet{Gear00} estimate
its redshift to lie between 2 and 4.5.

Lockman850.1 is the brightest SCUBA source detected in the
Lockman Hole ISOPHOT survey region with an extremely red
($K\sim 20$, $I-K>6.2$) and extended optical counterpart
\citep{Lutz01}.  Comparing the observed SED to that of Arp~220,
Lutz et al. estimate its redshift to be around 3, which agrees
well with our photometric redshift $z_{ph}=2.72$.

SMM~J14009+0252 is one of the submm sources identified in
the Abell~1835 field by \citet{Ivison00} with a faint 
optical counterpart ($K\sim 21$).  We derive $z_{ph}=1.30$,
but the SED fit shown in Figure~\ref{fig:4seds2}
for the dust spectrum is poor ($\chi^2_n = 2.3$).
Ivison et al. rejected their own similarly modest redshift 
estimate because this galaxy is 3 times more luminous than
HR~10 in the submm while it is more than 10 times fainter in $K$.
Its 1.4 GHz flux density is rather large for a SCUBA source,
indicating a possible presence of a radio AGN.  The
SED fit for the dust spectrum alone suggests $z\sim 3.5$.

FIRBACK~J1608+5418 is one of the FIR sources identified by the 
FIRBACK deep ISOPHOT survey \citep{Dole01}.  Incorporating its 
350 $\mu$m and 450 $\mu$m detection at 
the CSO and the VLA detection at 1.4 GHz, \citet{Benford} estimated 
its redshift to be around $z\sim 1.5$. Our photometric redshift of
$z_{ph}\sim0.8$ is substantially smaller.  
Its radio-submm flux density ratio suggests a presence of a radio AGN, 
and it is likely that higher than expected 
radio continuum flux density has led to a poor fit ($\chi^2_n = 17$).
Excluding the radio measurements, an SED
analysis for the submm/FIR dust peak alone suggests $z\sim 0.7$ 
with about 4 times larger FIR luminosity than Arp~220.

\bigskip
\section{Discussions and Summary \label{sec:summary}}

The above discussions of the SED modeling for the individual 
submm detected 
galaxies clearly indicate that systematic SED variation among the
individual galaxies is the dominant factor over the statistical
uncertainty associated with the SED measurements.  Here we discuss 
in detail the nature and the magnitude of the systematic effects 
associated with the dust and radio
emission as well as the SED sampling and the intrinsic calibration
uncertainties in the measurements. 
Their combined contribution to the photometric redshift 
uncertainty is quantified in \S~\ref{sec:err-total}.

\subsection{Effects of Dust Temperature and Emissivity
\label{sec:err-dust}}

Doppler shift of a dust spectrum to a higher redshift 
is exactly equivalent to lowering of dust 
temperature, and departure in $T_d$ from the template SED  
translates directly to a redshift error \citep[see][for further
discussions]{Blain99b}.  In other words, if the dust temperature
of a galaxy is actually higher than our template, then the resulting
$z_{ph}$ would be an under-estimate.  While inherently subject to
an object-to-object variation, the magnitude of uncertainty due to 
the spread in $T_d$ is well understood 
quantitatively: $\Delta z \sim {{\Delta T_d}\over{T_d}}(1+z)$.  
If $T_d$ is higher or lower by 10 K, the resulting 
error in the photometric redshift is about 0.3 at $z=1$, and it 
grows as $(1+z)$ reaching about 0.7 at $z=3$.  

Dust emissivity $\beta$ is as important as dust temperature $T_d$
because it determines the location of the low frequency edge 
of the thermal dust feature in the SEDs.   There is significant 
degeneracy between $\beta$ and $T_d$ \citep[e.g.][]{Dunne01},
and the decoupling the two becomes very difficult even when 
the dust spectrum is well sampled.  In practice only a few
measurements along the rising part of the dust SED are
measured, and any deficiency in $T_d$ is at least partly offset 
by the compensating effect of $\beta$.
A trend supporting this effect is seen in Figure~\ref{fig:betaTd}
as a broad trend of decreasing $\beta$ with increasing $T_d$.
A better demonstration of this effect is seen in Figure~\ref{fig:a220sed}
where the submm SED points alone might be fit well by 
a dust spectrum with $T_d\sim 45$ K and $\beta\sim 1.5$.
The observed trend of increasing $T_d$ and deceasing $\beta$
with increasing $SFR$ is also clearly seen in the theoretical modeling
of dust heating and emission using a self-consistent 3-D radiative
transfer code by \citet[][]{Misselt01}. 
Therefore the error estimate based on the $T_d$ alone might
serve only as a reasonable upper bound.  

\subsection{Effects of the Radio Spectrum
\label{sec:err-radio}}

Another reason why the photometric redshift error based on 
the scatter in $T_d$ alone might be an over-estimate is
that our SED fitting scheme relies as much on the 
spectral trough between the radio continuum and thermal
dust emission as the dust spectrum itself.  If this
spectral trough is considered as the primary SED feature from which
the photometric redshift information is derived, then it
is easy to see that the variation in the dust spectrum is 
strongly modulated by the radio spectrum which is
not affected.  

The radio spectrum 
has its own uncertainty in its overall scaling as shown by the
distribution of $f_{nth}$ (see Figure~\ref{fig:fnth}),
and its contribution to the photometric redshift error has
already been alluded to in \S~\ref{sec:tests}.
Presence of a powerful radio AGN was a severe limiting factor
for our earlier redshift estimation based on the radio-to-submm
spectral index \citep{Carilli99}, and it is not surprising
that radio AGNs pose the biggest obstacle for the SED fitting
photometric technique as well.  The photometric redshifts
derived for an ensemble of QSOs that are  
associated with dusty hosts are given at the bottom of
Table~\ref{tab:ztable} (also see Fig.~\ref{fig:comparez}).  
Although there are exceptions such
as LBQS~1230+1627 where $z_{ph} \sim z_{sp}$, the new photometric
method does not fare much better than the spectral index
method -- the median $\chi^2_n \sim 7$.  

In practice, radio AGNs are not likely to limit the
photometric redshift technique utilizing a starburst SED
template because only a small minority of the known submm
galaxies show evidence of hosting a powerful AGN.  
Comparisons of deep X-ray and submm surveys have shown
very little overlap between the detected sources 
\citep[e.g.][]{Fabian00,Horns00,Barger01}.
At least three out of five submm galaxies discussed in 
\S~\ref{sec:tests} are known to be an AGN host, but the presence of
an AGN also makes their spectroscopic redshift measurement possible
and the use of a photometric technique unnecessary.
The total fraction of infrared selected galaxies whose
radio-to-FIR SED shows a clear sign of energetic AGN is
only a few percent in the local universe \citep{Yun01},
and this fraction appears to remain roughly the same at higher
redshifts.

\subsection{SED Sampling and Calibration \label{sec:err-cal}}

Another important source of uncertainty is poor sampling of the SED.
Sampling only the radio part of the SED offers virtually no
redshift information, and the same is true if only the
submm part of the SED is measured.
On the other hand, because we are fitting the SED features changing 
over logarithmic scales, only a few well placed SED data points are 
needed to derived the redshifts \citep{Carilli99}. 
More than one measurement on both side of the SED trough is highly
desirable for a successful photometric redshift determination.

\begin{figure}
\plotone{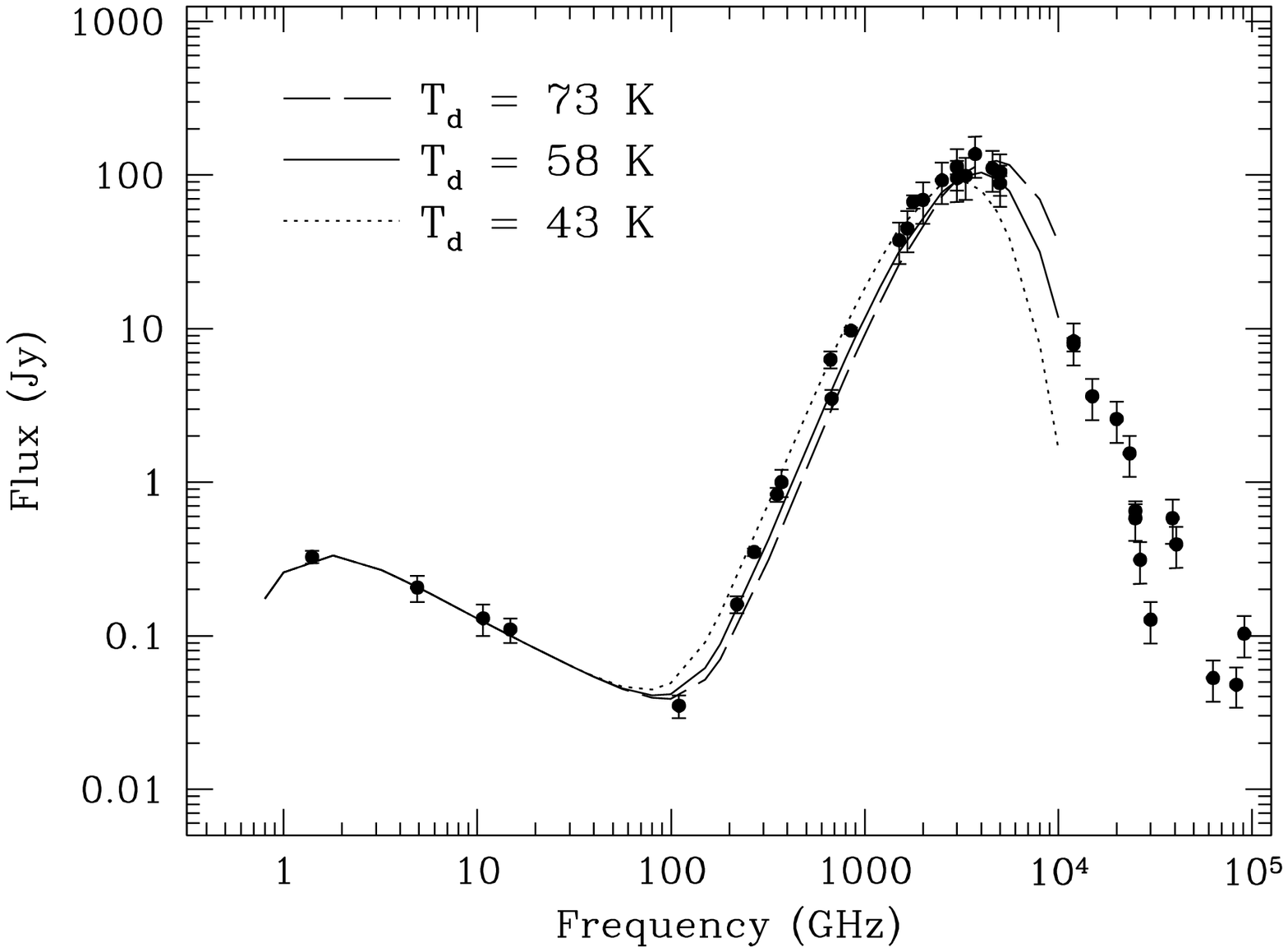}
\caption{Radio-to-IR SED for Arp~220 is shown along with the model 
starburst SEDs with $\beta=1.35$ and $T_d$ = 43 K, 58 K, and 73 K 
are shown.  The radio-FIR normalization term $f_{nth}$ is set to 1.0,
which is the Galactic normalization by Condon (1992). \label{fig:a220sed}}
\end{figure}

A related problem is the calibration and the relative weighting 
of individual SED measurements.  Measurement uncertainties
found in the literature often do not fully account for the overall
calibration accuracy as demonstrated by the magnitude of the scatter
and the sizes of the error bars plotted for the SED measurements
of Arp~220 in Figure~\ref{fig:a220sed}.  For example, the IRAS
60 $\mu$m and 100 $\mu$m measurements found in the Point Source
Catalog (PSC) and the Faint Source Catalog (FSC) are known to
have systemic calibration differences of order 10\% while
the flux densities reported in the FSC often include uncertainties
of only a few percent.  For deriving the SED template from the
23 infrared selected starbursts, we adopted flux densities
in the FSC but increased measurement uncertainties
to 10\% (unless the reported uncertainty was larger).
This re-calibration of the measurement uncertainty is more than
cosmetic since the relative weight of the data points are
directly reflected in the $\chi^2$ minimization.  We found this 
re-weighting to be critically important in 
utilizing the submm measurements.  Similarly, all radio
continuum measurements are assigned at least 10\% uncertainty
in order to account for the overall uncertainty in the flux calibration.
No effort was made to re-calibrate any of the submm data
points since most measurements generally carry relatively large
fractional uncertainties, but some are clearly under-estimates
based on the SED plots such as shown in Figure~\ref{fig:a220sed}.
The same problems also plague the reported SED measurements 
of many submm galaxies, and they will inevitably impact the
accuracy of the photometric redshifts derived from these data.
One way to improve the situation in the future would be employing
a frequency selective bolometer \citep[e.g. ][]{Kowitt96,Meyer01} that
can make simultaneous measurements of several submm bands
with accurate relative calibration between the measurement bands.

\subsection{Estimate of Uncertainty in the Radio-to-FIR SED Technique
\label{sec:err-total}}

The uncertainty of $\sigma_{z}\sim 0.3(1+z)$ derived from the
scatter in dust temperature is probably the upper bound to the
error one may expect from the photometric technique using the
dusty starburst SED template.  The compensating effects of the
dust emissivity and the radio continuum are more difficult to
quantify as are the uncertainties contributed by the scatter
in $f_{nth}$.  One way to estimate
the collective uncertainty of the SED photometric technique
is to apply this technique to the same 23 dusty starbursts
from which the SED template was derived.  This is not entirely
circuitous since our SED template is based only on the average
properties of these galaxies and has no knowledge of dispersions.  
At the least, we may be able to confirm the impact of the
10 K scatter in the dust temperature on the dispersion
in $z_{ph}$ if dust temperature variation dominates the uncertainty
in determining photometric redshifts.

The resulting ``photometric redshift'' $z_{ph}$ for the 23 dusty 
starbursts are listed in the last column of Table~\ref{tab:23gals}.
There are several galaxies with negative $z_{ph}$ since
the impact of higher dust temperature 
for a $z=0$ galaxy with $T_d > 58$ K would result in $z_{ph}<0$.  The
$\chi^2$ minimization program is modified to search a redshift
range of $-1\ge z_{ph} \ge +1$ with modified scaling along
the flux density axis in order to remove the non-physical 
impact of negative redshift on $D_L$.  A handful of galaxies
with negative redshift are known (e.g. M81), but negative 
redshift is generally considered non-physical.  Here, the only
physically meaningful interpretation of
a negative $z_{ph}$ we derive is the measure of
the magnitude of departure in the radio and dust property
from the template SED, just the same way any uncertainty in
$z_{ph}$ should be interpreted at any redshift.

\begin{figure}
\epsscale{0.75}
\plotone{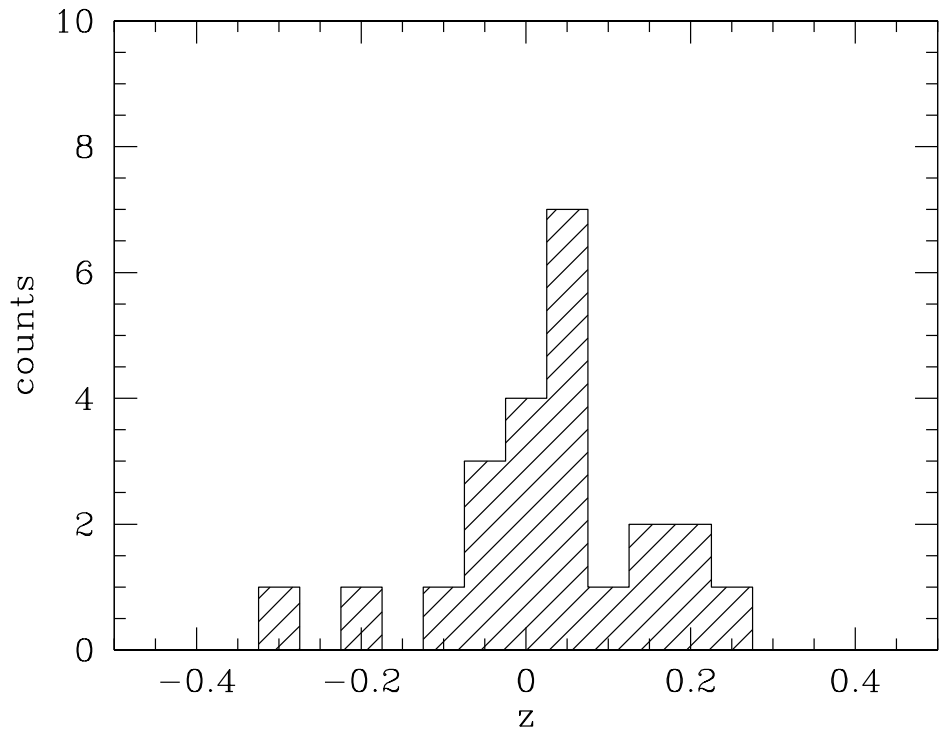}
\caption{Histogram of ``redshifts'' for the 23 luminous infrared
starbursts derived using our SED template.
\label{fig:z0histo}}
\end{figure}

A histogram of the resulting photometric
redshifts shown in Figure~\ref{fig:z0histo} suggests that the
median ``redshift'' is about +0.05, suggesting the template
SED may be slightly biased, but this offset is probably
not very significant given the range of ``redshifts'' derived.
``Photometric redshifts'' as large as +0.3 and as small as $-$0.3 are 
found as expected from the scatter in $T_d$.  
The two extreme negative ``$z_{ph}$'' object IRAS~08572+3915 and 
Mrk~231 have $T_d$ of 74 K and 72 K, respectively, following 
the general trend expected of the dust temperature variation.  On the 
other hand, two other dusty starbursts with characteristic $T_d\ge70$ K,
IRAS 15250+3609 and IRAS~05189$-$2524, have ``$z_{ph}$'' of +0.01
and +0.05, clearly demonstrating that there are other compensating
effects and that the scatter in $T_d$ alone does not dictate
the overall photometric redshift uncertainty.  

Regardless of the underlying causes, the ``photometric redshifts''
for 2/3 of all galaxies lie within $\Delta z \le 0.10$, and
we estimate the collective uncertainty of the photometric redshift
technique, including the variations in the radio and dust
properties as well as the uncertainties in the $\chi^2$ minimization
process, is about 0.1 as long as this technique is applied to
luminous dusty starburst galaxies only.  Allowing for the ($1+z$) frequency
folding of the Doppler effect, we estimate an overall 
uncertainty of $\sigma_z \sim 0.1 (1+z)$ with an upper bound
in redshift uncertainty of about $0.3(1+z)$.

In either case, this photometric redshift technique utilizing
the radio-to-FIR dusty starburst SED represents a significant
step forward, particularly at high redshifts, 
when compared with existing methods.  The full
potential of this method will be realized when the sources
identified by
several large deep, multi-frequency surveys planned in the
immediate future (e.g. SIRTF Legacy Surveys) are analyzed together
to reveal an accurate redshift distribution of luminous
dusty galaxies at high redshift.  
\bigskip

\acknowledgements

The National Radio Astronomy Observatory is a facility of the 
National Science Foundation
operated under cooperative agreement by Associated Universities, Inc.
Some of the data presented here are obtained from the
NASA/IPAC Extragalactic Database (NED), which is operated by the Jet
Propulsion Laboratory, California Institute of Technology, under
contract with the National Aeronautical and Space Administration.

\clearpage

\clearpage

\clearpage

\clearpage

\clearpage

\clearpage

\clearpage

\clearpage

\begin{deluxetable}{lcccccc}
\tablewidth{32pc}
%\tablewidth{26pc}
%\tabletypesize{\scriptsize}
\tablecaption{Spectral Energy Distribution Model Fits \label{tab:23gals}}
\tablehead{
\colhead{Name} & \colhead{L(FIR)} & \colhead{$SFR$\tablenotemark{a}} & 
\colhead{$T_d$} & \colhead{$\beta$\tablenotemark{b}} & 
\colhead{$f_{nth}$\tablenotemark{c}} & \colhead{``$z_{ph}$''} \\
 & ($10^{10}L_\odot$) & ($M_\odot~yr^{-1}$) & (K) & & & }
\startdata
Zw049.057 & 10  & 18  & 54 & 1.30 & 0.6 & +0.06 \\
NGC~5256  & 11  & 19  & 51 & 1.35 & 4.0 & +0.07 \\
MCG+00-29-023  & 11 & 19  & 49 & 1.15 & 1.3 & +0.25 \\
NGC~3110  & 12 & 20  & 46 & 1.35 & 1.2 & +0.19 \\
Mrk~331   & 16  & 27  & 57 & 1.15 & 0.8 & +0.06 \\
NGC~1614  & 21  & 37  & 61 & 1.40 & 1.0 & $-$0.07 \\
UGC~2369  & 21  & 37  & 52 & 1.30 & 1.2 & +0.16 \\
NGC~2623  & 21  & 37  & 57 & 1.45 & 1.1 & $-$0.07 \\
Arp~193   & 25  & 42  & 47 & 1.70 & 1.5 & +0.01 \\
Arp~148   & 25  & 42  & 50 & 1.20 & 0.9 & +0.22 \\
IRAS~10173+0828 & 33  & 57  & 69 & 1.10 & 0.6 & +0.02 \\
NGC~6240  & 35  & 61  & 49 & 1.45 & 3.9 & $-$0.07 \\
Mrk~848   & 35  & 61  & 62 & 1.20 & 1.6 & +0.06 \\
IRAS~15250+3609 & 48 & 82 & 74 & 1.05 & 1.1 & +0.01 \\
UGC~5101  & 54  & 93  & 47 & 1.40 & 2.5 & +0.14 \\
IRAS~08572+3915 & 55 & 93 & 74 & 1.60 & 0.3 & $-$0.22 \\
IRAS~05189$-$2524 & 57 & 98 & 70 & 1.05 & 0.6 & +0.05 \\
IRAS~10565+24 & 60  & 104  & 55 & 1.55 & 0.4 & +0.06 \\
Mrk~273   & 73  & 127 & 61 & 1.40 & 1.7 & $-$0.10 \\
Arp~220   & 95  & 163 & 59 & 1.15 & 1.0 & +0.05 \\
IRAS~12112+0305 & 106 & 183 & 63 & 1.25 & 0.8 & +0.02 \\
IRAS~14348$-$1447 & 109 & 187 & 63 & 1.50 & 1.0 & +0.09 \\
Mrk~231   & 137 & 236 & 72 & 1.35 & 2.0 & $-$0.30 \\
\tablenotetext{a}{$SFR=L_{FIR}/(5.8\times10^9L_\odot)~M_\odot
~{\rm yr}^{-1}$ \citep{Kennicutt98}}
\tablenotetext{b}{Dust emissivity.}
\tablenotetext{c}{Multiplicative correction factor for  
non-thermal radio continuum emission.  The $f_{nth}=1.0$ corresponds 
to the Galactic normalization adopted by Condon (1992).}
\enddata
\end{deluxetable}

\clearpage

\begin{deluxetable}{lccccccl}
\tablewidth{33pc}
\tabletypesize{\scriptsize}
\tablecaption{Redshift Estimates for Submm Sources \label{tab:ztable}}
\tablehead{
\colhead{Name} & \colhead{$z_{sp}$} & \colhead{$z_{SI}$} & 
\colhead{$z_{ph}$} &\colhead{$SFR$\tablenotemark{a}} &
\colhead{$N_{tot}~(N_{det})$\tablenotemark{b}} & 
\colhead{$\chi^2_n$\tablenotemark{c}} & \colhead{Refs.} }
\startdata
\sidehead{Known Redshifts:}
CUDSS14.18 & 0.66 & $1.12_{-0.45}^{+0.53}$ & 
$1.11\pm0.21$ & 200 & 5 (3) & 0.02 & 1 \\
SMM J02399$-$0134 & 1.06 & $1.17_{-0.45}^{+0.53}$ & 
$0.93\pm0.19$ & 525 & 5 (3) & 0.5 & 2 \\
HR10 & 1.44 & $1.49_{-0.53}^{+0.75}$ & 
$1.75\pm0.28$ & 380 & 7 (5) & 0.7 & 3,4,5 \\
SMM J14011+0252 & 2.57 & $2.53_{-0.82}^{+1.24}$ & 
$2.73\pm0.37$ & 850 & 5 (4) & 1.0 & 2,6 \\
SMM J02399$-$0136 & 2.80 & $1.65_{-0.53}^{+0.80}$ & 
$2.83\pm0.38$ & 1400 & 9 (5) & 28 & 2,7 \\
\sidehead{No Known Redshifts:}
FIRBACK J1608+5418 & -- & $0.05_{-0.05}^{+0.21}$ & 
$0.84\pm0.18$ & 690 & 7 (5) & 17 & 12,13,14 \\
SMM J14009+0252 & -- & $1.27_{-0.54}^{+0.59}$ & 
$1.30\pm0.23$ & 980 & 5 (4) & 2.3 & 2,6 \\
CUDSS14.1	& -- & $2.01_{-0.71}^{+1.10}$ & 
$2.06\pm0.31$ & 520 & 7 (4) & 0.5 & 10,11 \\
Lockman850.1	& -- & $2.95_{-0.98}^{+1.49}$ & 
$2.72\pm0.37$ & 570 & 8 (5) & 0.2 & 15 \\
SMM J00266+1708 & -- & $3.49_{-1.23}^{+2.03}$ & 
$3.50\pm0.45$ & 1200 & 6 (4) & 0.2 & 2 \\
SMM J09429+4658 & -- & $\ge 3.6$ & 
$3.86\pm0.49$ & 890 & 4 (4) & 4.0 & 2,6 \\
HDF850.1	& -- & $>2.6$ & $4.11\pm0.51$ & 380 & 8 (4) & 0.8 & 8,9 \\
\sidehead{QSOs:}
FSC 10214+4724 & 2.28 & $1.49_{-0.51}^{+0.74}$ & $ 1.51\pm0.25$ & 
2600 & 9 (9) & 7.6 & 16,17 \\
H 1413+117    & 2.56 & $0.88_{-0.34}^{+0.41}$ & $ 0.31\pm0.13$ & 
850 & 9 (9) & 28 & 17,18 \\
LBQS 1230+1627 & 2.74 & $2.19_{-0.75}^{+1.16}$ & $ 2.66\pm0.37$ & 
1300 & 4 (4) & 0.5 & 19,20,21 \\
SMM J04135+1027 & 2.83 & $\ge 0.9$ & $1.83\pm0.28$ & 1560 & 4 (3) & 0.8 & 22 \\
APM 08279+5255  & 3.87 & $2.18_{-0.71}^{+1.10}$ & $ 1.74\pm0.28$ &
4900 & 7 (7) & 9.4 & 23,24,25 \\
BR 1335$-$0417  & 4.40 & $1.81_{-0.62}^{+0.95}$ & $ 2.84\pm0.38$ & 
1200 & 5 (5) & 7.0 & 17,19,20,21,26 \\
BR 0952$-$0115  & 4.43 & $1.35_{-0.46}^{+0.63}$ & 
$ 0.96\pm0.20$ & 460 & 5 (4) & 5.6 & 19,20,21,26 \\
BR 1202$-$0725  & 4.70 & $3.13_{-1.14}^{+1.81}$ & $ 3.20\pm0.42$ & 
2300 & 11 (9) & 3.1 & 17,19,20,21,26 \\
\enddata
\tablenotetext{a}{$SFR$ is in $M_\odot$ yr$^{-1}$.}
\tablenotetext{b}{$N_{tot}$ and $N_{det}$ are total number of SED data
used for the fit and the number of detections (i.e. $N_{tot}=N_{det}
+ N_{limits}$).}
\tablenotetext{c}{$\chi^2_n \equiv \chi^2/N_{det}$.}
\tablerefs{
(1) Eales et al. (2001); (2) Smail et al. (2000); (3) Cimatti et al. (1998);
(4) Dey et al. (1999); (5) Andreani et al. (2000); (6) Ivison et al. (2000);
(7) Ivison et al. (1998); (8) Hughes et al. (1998); (9) Downes et al. (1999a);
(10) Eales et al. (2000); (11) Gear et al. (2000); (12) Benford (1999);
(13) Scott et al. (2000); (14) Dole et al. (2001); (15) Lutz et al. (2001);
(16) Rowan-Robinson et al. (1993); (17) Benford et al. (1999); 
(18) Barvainis et al. (1992); (19) Omont et al. (1996); 
(20) Guilloteau et al. (1999); (21) Yun et al. (2000); 
(22) Knudsen et al. (2000); (23) Irwin et al. (1998); 
(24) Lewis et al. (1998); (25) Downes et al. (1999b); 
(26) McMahon et al. (1999)}
\end{deluxetable}

\end{document}